\documentclass{elsart}%
\usepackage{makeidx}
\usepackage{amsmath}
\usepackage{amsfonts}
\usepackage{amssymb}
\usepackage{graphicx}
\usepackage{lineno}%
\begin{document}
\begin{frontmatter}
\title{Statistics of weighted Poisson events and its applications}
\author{G. Bohm}
\address{Deutsches Elektronen-Synchrotron DESY, D-15738 Zeuthen, Germany} and
\author{G. Zech\thanksref{mail}}
\thanks[mail]{Corresponding author, email: zech@physik.uni-siegen.de}
\address{Universit\"at Siegen, D-57068 Siegen, Germany}
\begin{abstract}
The statistics of the sum of random weights where the number of weights is Poisson
distributed has important applications in nuclear physics, particle physics and astrophysics.
Events are frequently weighted according to their acceptance or relevance to a certain type of reaction.
The sum is described by the compound Poisson distribution (CPD) which is shortly reviewed. It is shown that
the CPD can be approximated by a scaled Poisson distribution (SPD). The SPD is applied to
parameter estimation in situations where the data are distorted by resolution effects.
It performs considerably better than the normal approximation that is usually used.
A special Poisson bootstrap technique is presented which permits to derive confidence limits
for observations following the CPD.
\end{abstract}
\end{frontmatter}

\section{Introduction}

In the analysis of the data collected in particle experiments, frequently
weighted events have to be dealt with. For instance, losses due to a limited
acceptance of the detector are \ corrected by weighting each events by the
inverse of its detection probability. The sum of the weights is used to
estimate the number of incident particles. Frequently an event cannot be
uniquely assigned to a signal or to a background and is attributed a weight.
It is necessary to associate confidence limits, upper or lower limits to the
sum of these weights and we want to apply goodness-of-fit tests to histograms
of weighted events. When data are compared \ to different theoretical
predictions, weighting simplifies the computation and sometimes it is
unavoidable. Of special importance is parameter estimation from histograms
where the measured variable is distorted by the limited resolution and
acceptance of the detector. To evaluate these effects, Monte Carlo simulations
have to be performed. In the simulation a p.d.f. $f_{0}(x)$ is assumed to
describe the data where $x$ represents the set of event variables that are
measured. The simulation often requires considerable computer power. To change
the p.d.f. of the simulation to $f(x)$, the simulated events are weighted by
\ $f(x)/f_{0}(x)$. This is unavoidable if $f(x|\theta)$ depends on one or
several parameters $\theta$ that have to be estimated. During the fitting
procedure where the experimental distributions are compared to the simulated
ones, the parameters and correspondingly the weights $f(x|\theta
)/f(x|\theta_{0})$ are varied.

The distribution of the sum $x$ of a Poisson distributed number of weights is
described by a \emph{compound Poisson distribution }(CPD), provided the
weights can be considered as independent and identical distributed random
variables. This condition is realized in the majority of experimental
situations. The CPD applies to a large number of problems. In astrophysics,
for instance, the total energy of random airshowers follows a CPD. Also
outside physics the CPD is widely used to model processes like the sum of
claims in car accidents and other insurance cases which are assumed to occur
randomly with different severity.

Usually we do not dispose of the distribution of the weights but have to base
the analysis on the observed weights of a sample of events. The mean value and
the variance of the weighted sum can be inferred directly from the
corresponding empirical values, but in many cases a more detailed knowledge of
the distribution is necessary. To deal with these situations, the
approximation of the CPD by a scaled Poisson distribution and a special
bootstrap technique are proposed.

In the first part of this article some properties of the CPD and an
approximation of it that is useful in the analysis of experimental data are
discussed. The second part contains applications where observed samples of
weights have to be analyzed and where the underlying distribution of the
weights is unknown. Parameter estimation with weighted Monte Carlo events
using a scaled Poisson distribution and the estimation of confidence limits
with Poisson bootstrap are studied.

\section{The compound Poisson distribution}

\subsection{Distribution of a sum of weighted Poisson numbers}

The distribution of a weighted Poisson number $x=wm$ with $m\sim
\mathcal{P}_{\lambda}(m)=e^{-\lambda}\lambda^{m}/m!$, $m=0,1,2,\ldots$ and the
weight $w$, a real valued positive parameter, $w>0$ is
\[
W(x)=\frac{e^{-\lambda}\lambda^{x/w}}{(x/w)!}\;.
\]

To evaluate the moments of the Poisson distribution, it is convenient to use
the cumulants. The moments $\mu_{k}$ of a distribution are polynomials of the
cumulants $\kappa_{1,}...,\kappa_{k}$ of the distribution. For the first two
moments $\mu,\sigma^{2}$, the skewness $\gamma_{1}$ and the excess $\gamma
_{2}$ the relations are $\mu=\kappa_{1}$, $\sigma^{2}=\kappa_{2}$, $\gamma
_{1}=\kappa_{3}/\kappa_{2}^{3/2}$, $\gamma_{2}=\kappa_{4}/\kappa_{2}^{2}$. The
cumulants $\kappa_{k}$ of the Poisson distribution $\mathcal{P}_{\lambda}$ are
especially simple, they are all identical and equal to the mean value
$\lambda$ and thus $\gamma_{1}=1/\lambda^{1/2}$ and $\gamma_{2}=1/\lambda$.
From the homogeneity of the cumulants follows for the cumulant $\kappa_{k}(x)$
of order $k$ of the distribution of $x$ the relation $\kappa_{k}(x)=\kappa
_{k}(wm)=w^{k}\lambda$.

We consider now two Poisson processes with random variables $n_{1}$ and
$n_{2}$ and mean values $\lambda_{1}$ and $\lambda_{2}$. We are interested in
the distribution of the weighted sum $x=x_{1}+x_{2}=w_{1}n_{1}+w_{2}n_{2}$
with positive weights $w_{1}$, $w_{2}$.

The mean value and the variance of $x$ are%
\begin{align}
\mathrm{E}(x)  &  =w_{1}\lambda_{1}+w_{2}\lambda_{2}\;,\label{mean2}\\
\mathrm{Var}(x)  &  =w_{1}^{2}\lambda_{1}+w_{2}^{2}\lambda_{2}\;. \label{var2}%
\end{align}

These results follow from the properties of expected values and are
intuitively clear. The cumulant of the distribution of the sum of two
independent random variables $x_{1}$ and $x_{2}$ is the sum of the two
cumulants:
\begin{equation}
\kappa_{k}=w_{1}^{k}\lambda_{1}+w_{2}^{k}\lambda_{2}\;. \label{cum2}%
\end{equation}
Relation (\ref{cum2}) can be generalized to $N$ Poisson processes with mean
values $\lambda_{i}$:%
\begin{equation}
\kappa_{k}=%
{\displaystyle\sum\limits_{i=1}^{N}}
w_{i}^{k}\lambda_{i}\;. \label{varN}%
\end{equation}

We will see below that the case is of special interest where all mean values
$\lambda_{i}$ are equal. With $\lambda_{i}=\lambda/N$, and $x=\Sigma w_{i}$
the modified relation for the cumulants is%
\begin{equation}
\kappa_{k}=\lambda\Sigma w_{i}^{k}/N=\lambda\left\langle w^{k}\right\rangle
\;, \label{cummu}%
\end{equation}
and $\mu$, $\sigma$, $\gamma_{1}$,$\gamma_{2}$ are:%
\begin{align}
\mu &  =\lambda\Sigma_{i}w_{i}/N=\lambda\left\langle w\right\rangle
\;,\label{mu}\\
\sigma^{2}  &  =\lambda\Sigma_{i}w_{i}^{2}/N=\lambda\left\langle
w^{2}\right\rangle \;,\label{sigma2}\\
\gamma_{1}  &  =\frac{\lambda\Sigma_{i}w_{i}^{3}/N}{\sigma^{3}}=\frac
{\left\langle w^{3}\right\rangle }{\lambda^{1/2}\left\langle w^{2}%
\right\rangle ^{3/2}}\;,\label{gamma1}\\
\gamma_{2}  &  =\frac{\lambda\Sigma_{i}w_{i}^{4}/N}{\sigma^{4}}=\frac
{\left\langle w^{4}\right\rangle }{\lambda\left\langle w^{2}\right\rangle
^{2}}\;. \label{gamma2}%
\end{align}
Here $\left\langle v\right\rangle $ denotes the mean value $\Sigma_{i=1}%
^{N}v_{i}/N$.

So far we have treated the weights as parameters, while according to the
definition of a compound Poisson process, the weights are random variables.

\subsection{Distribution of the sum of random weights}

In most applications of particle physics the distribution of the sum of
individually weighted events is of interest. The number $n$ of events is
described by a Poisson distribution and to each event a random weight is
associated. Instead of the $N$ independent Poisson processes with mean values
$\lambda_{i}$ and random variables $n_{i}$ we can consider the random variable
$n=\Sigma n_{i}$ as the result of a single Poisson process with $\lambda
=\Sigma\lambda_{i}$. The numbers $n_{i}$ are then chosen from a multinomial
distribution where $n$ is distributed to the $N$ different weight classes with
probabilities $\varepsilon_{i}=\lambda_{i}/\lambda$, i.e. a weight $w_{i}$ is
chosen with probability $\varepsilon_{i}$:
\begin{align}
W(n_{1},...,n_{N}) &  =%
{\displaystyle\prod_{i=1}^{N}}
\mathcal{P}(n_{i}|\lambda_{i})=\mathcal{P}_{\lambda}(n)\mathcal{M}%
_{\varepsilon_{1},...,\varepsilon_{N}}^{n}(n_{1},...,n_{N}%
)\;,\label{poissonmulti}\\
\mathcal{M}_{\varepsilon_{1},...,\varepsilon_{N}}^{n}(n_{1},...,n_{N}) &  =n!%
{\displaystyle\prod\limits_{i=1}^{N}}
\varepsilon_{i}^{n_{i}}\left/
{\displaystyle\prod\limits_{i=1}^{N}}
n_{i}!\right.  \;.\label{poissonmulti1}%
\end{align}

The validity of (\ref{poissonmulti}) is seen from the following identity for
the binomial case,%
\[
\mathcal{P}_{\lambda}\mathcal{M}_{\lambda_{1}/\lambda,\lambda_{2}/\lambda}%
^{n}=\frac{e^{-\lambda}\lambda^{n}}{n!}\frac{n!}{n_{1}!n_{2}!}\frac
{\lambda_{1}^{n_{1}}\lambda_{2}^{n_{2}}}{\lambda^{n_{1}}\lambda^{n_{2}}}%
=\frac{e^{-(\lambda_{1}+\lambda_{2})}\lambda_{1}^{n_{1}}\lambda_{2}^{n_{2}}%
}{n_{1}!n_{2}!}=\mathcal{P}_{\lambda_{1}}\mathcal{P}_{\lambda_{2}}\;
\]
which is easily generalized to the multinomial case.

It does not matter whether we describe the distribution of $x$ by independent
Poisson distributions or by the product of a single Poisson distribution with
a multinomial distribution. If all probabilities are equal, $\varepsilon
_{i}=1/N$, the multinomial distribution describes a random selection of the
weights $w_{i}$ out of the $N$ weights with equal probabilities $1/N$. The
formulas (\ref{mu}) to (\ref{gamma2}) remain valid.

To describe a continuous weight distribution $f(w)$ with finite variance, the
limit $N\rightarrow\infty$ has to be considered. Again our formulas remain
valid with $\varepsilon N=1$. We get $x=\Sigma_{i=1}^{n}w_{i}$. The mean
values $\left\langle w^{k}\right\rangle $ in (\ref{cummu}), (\ref{mu}),
(\ref{sigma2}), (\ref{gamma1}), (\ref{gamma2}) are to be replaced by the
corresponding expected values $\mathrm{E}(w^{k})$, e.g. the moments of the
weight distribution.

\section{Approximation by a scaled Poisson distribution}

To analyze the sum of weights of observed samples where the underlying weight
distribution is not known, it is necessary to approximate the CPD. According
to the central limit theorem, the sum of weighted Poisson random numbers with
mean number $\lambda$ and expected weight $\mathrm{E}(w)$ can asymptotically,
for $\lambda\rightarrow\infty$, be described by a normal distribution with
mean $\mu=$ $\lambda\mathrm{E}(w)$ and variance $\sigma^{2}=\lambda
\mathrm{E}(w^{2})$, provided the expected values exist. The speed of
convergence with $\lambda$ depends on the distribution of the weights.

As is demonstrated below, the moments of the CPD are closer to those of a
\emph{scaled Poisson distribution} (SPD) than to the moments of the normal
distribution. Especially, if the weight distribution is narrow, the SPD is a
very good approximation of the CPD and in the limit where all weights are
identical, it coincides with the CPD.

The SPD is fixed by the requirement that the first two moments of the CPD have
to be reproduced. We define an equivalent mean value $\tilde{\lambda}$,
\begin{align}
\tilde{\lambda}  &  =\lambda\frac{\mathrm{E}(w)^{2}}{\mathrm{E}(w^{2}%
)}\label{enoe}\\
&  =\mu\frac{\mathrm{E}(w)}{\mathrm{E}(w^{2})}=\frac{\mu}{s}\;,
\end{align}
an equivalent random variable $\tilde{n}\sim\mathcal{P}_{\tilde{\lambda}}$, a
scale factor $s$,%
\begin{equation}
s=\frac{\mathrm{E}(w^{2})}{\mathrm{E}(w)}\;, \label{scale}%
\end{equation}
and a scaled random variable $\tilde{x}=s\tilde{n}$ such that the expected
value $\mathrm{E}(\tilde{x})=\mathrm{E}(x)=\mu$ and the variance
$\mathrm{Var}(\tilde{x})=\mathrm{Var}(x)=\sigma^{2}$. The cumulants of the
scaled distribution are $\tilde{\kappa}_{k}=s^{k}\tilde{\lambda}$.

To evaluate the quality of the approximation of the CPD by the SPD, we compare
the cumulants of the two distributions and form the ratios $\kappa_{k}%
/\tilde{\kappa}_{k}$. Per definition the ratios for $k=1,2$ agree because the
two lowest moments agree.

The skewness and excess for the two distributions in terms of the expected
values of powers $k$ of $w$, $\mathrm{E}(w^{k})$ are according to
(\ref{gamma1}), (\ref{gamma2}) and (\ref{enoe}):%
\begin{align}
\gamma_{1}  &  =\frac{\mathrm{E}(w^{3})}{\lambda^{1/2}\mathrm{E}(w^{2})^{3/2}%
}\;,\label{g1}\\
\gamma_{2}  &  =\frac{\mathrm{E}(w^{4})}{\lambda\mathrm{E}(w^{2})^{2}%
}\;,\label{g2}\\
\tilde{\gamma}_{1}  &  =\frac{1}{\tilde{\lambda}^{1/2}}=\left[  \frac
{\mathrm{E}(w^{2})}{\lambda\mathrm{E}(w)^{2}}\right]  ^{1/2}\;,
\label{g1tilde}\\
\tilde{\gamma}_{2}  &  =\frac{1}{\tilde{\lambda}}=\frac{\mathrm{E}(w^{2}%
)}{\lambda\mathrm{E}(w)^{2}}\;. \label{g2tilde}%
\end{align}

For the ratios we obtain%
\begin{align}
\frac{\gamma_{1}}{\tilde{\gamma}_{1}}  &  =\frac{\mathrm{E}(w^{3}%
)\mathrm{E}(w)}{\mathrm{E}(w^{2})^{2}}\geq1\;,\label{rg1}\\
\frac{\gamma_{2}}{\tilde{\gamma}_{2}}  &  =\frac{\mathrm{E}(w^{4}%
)\mathrm{E}(w)^{2}}{\mathrm{E}(w^{2})^{3}}\geq1\;. \label{rg2}%
\end{align}

The proof of these inequalities is given in the Appendix. As $\gamma_{1}$ and
$\gamma_{2}$ of the normal distribution are zero, the values $\gamma_{1}$ and
$\gamma_{2}$ of the CPD are closer to those of the SPD than to those of the
normal distribution. This property suggests that the SPD is a better
approximation to the CPD than the normal distribution. According to the
central limit theorem, CPD and SPD approach the normal distribution with
increasing $\tilde{\lambda}$. The equalities in (\ref{rg1}) and (\ref{rg2})
hold if all weights are equal. Remark that the ratios do not depend on the
expected number $\lambda$ of weights, only the moments of the weight
distribution enter. The ratios are close to unity in most practical cases.
They can become large if the weight distribution comprises weights that differ
considerably and especially if many small weights are combined with few large
weights. This is the case, for instance, for an exponential weight distribution.%

\begin{figure}
[ptb]
\begin{center}
\includegraphics[
trim=0.000000in 0.130982in 0.000000in 0.098694in,
height=4.1644in,
width=5.8165in
]%
{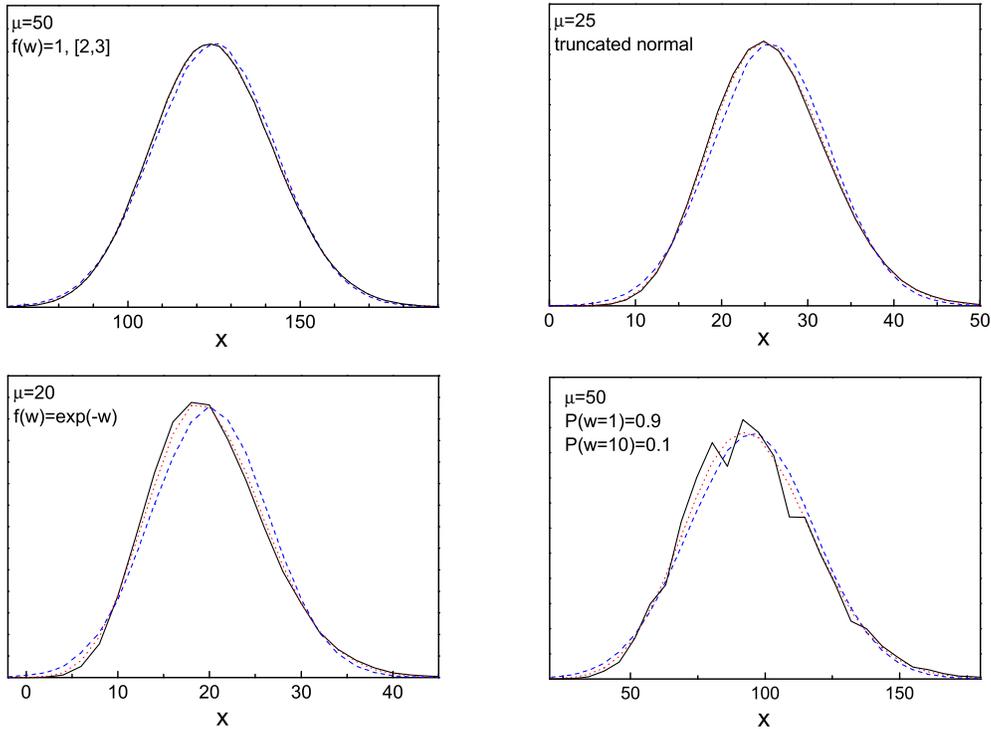}%
\caption{{\small Comparison of a CPD (solid) with a SPD (dotted) and a normal
distribution (dashed)}}%
\label{distcompare}%
\end{center}
\end{figure}

In Figure 1 the results of a simulation of CPDs with different weight
distributions is presented. The simulated events are collected into histogram
bins but the histograms are displayed as line graphs which are easier to read
than column graphs. Corresponding SPD distributions are generated with the
parameters chosen according to the relations (\ref{enoe}) and (\ref{scale}).
They are indicated by dotted lines. The approximations by normal distributions
are shown as dashed lines. Due to the discrete Poisson distribution the
histograms for the composite Poisson distribution and the SPD have pronounced
structures that makes it difficult to compare the results. To avoid at least
partially this disturbing effect, the binning was adapted to the steps of the
SPD. The weight distribution of the top left graph is uniform in the interval
$[2,3]$ and the weight distribution of the top right graph is a truncated,
renormalized normal distribution $\mathcal{N}_{t}(x|1,1)=c\mathcal{N}(x|1,1)$,
$x>0$ with mean and variance equal to $1$ where negative values are cut. In
both cases the approximation by the SPD is hardly distinguishable from the
CPD. In the bottom left graph the weights are exponentially distributed. This
case inhibits large weights with low frequency where the approximation by the
SPD is less good. Still it models the CPD reasonably well. In the bottom right
\ graph the weight distribution is discrete with the weight $w_{1}=1$ chosen
with probability $0.9$ and the weight $w_{2}=10$ chosen with probability
$0.1$. This is again an extreme situation. The SPD and the CPD agree
reasonably well globally, but have different discrete structures which result
in jumps caused by the binning. The examples show, that the approximation by
the SPD is mostly close to the CPD and that it is always superior to the
approximation by the normal distribution.

In Table 1 we compare skewness $\gamma_{1}$ and excess $\gamma_{2}$ of the SPD
to the values of the CPD. The mean values from $1000000$ simulated experiments
are taken. The mean number of weights is always $50$, e.g. $n\sim
\mathcal{P}_{50}(n)$. The weights used to obtain the first $3$ rows are
uniformly distributed in the indicated interval. The weights of the following
row are distributed according to $\exp(-w)$, the weights of the next row
follow the truncated normal distribution. The last two rows correspond to two
discrete weights, $w_{1}=$ $1$ and $w_{2}=10$ chosen with equal probabilities
and with $w_{1}=0.8$ and $w_{2}=0.2$, respectively. The second column
indicates the number of equivalent events $\tilde{\lambda}$ defined in
(\ref{enoe}), e.g. the number of unweighted events with the same relative
uncertainty as the weighted sum $x$. For example, the relative fluctuation
$\delta x/x$ of the sum $x$ of $n\sim\mathcal{P}_{50}(n)$ random weights with
$w\sim\exp(-w)$ is $1/\sqrt{25}=0.2$. The following columns contain the values
of $\gamma_{1}$ and $\gamma_{2}$ of the CPD and those from the scaled Poisson
distribution. The values of the normal approximation are $\gamma_{1}%
=\gamma_{2}=0$. The first two moments are per definition equal for the CPD and
the SPD.

\begin{table}[ptb]
\caption{Skewness and excess of the SPD approximation}%
\label{g1g2spd}
\begin{center}%
\begin{tabular}
[c]{|l|l|l|l|l|l|}\hline
type of weight & $\tilde{\lambda}$ & $\gamma_{1}$ & $\gamma_{2}$ &
$\tilde{\gamma}_{1}$ & $\tilde{\gamma}_{2}$\\\hline
$u[0,1]$ & $37.50$ & $0.184$ & $0.036$ & $0.163$ & $0.027$\\
$u[1,2]$ & $48.21$ & $0.149$ & $0.023$ & $0.144$ & $0.021$\\
$u[2,3]$ & $49.34$ & $0.144$ & $0.021$ & $0.142$ & $0.020$\\
$exp(-w)$ & $25.00$ & $0.300$ & $0.120$ & $0.200$ & $0.040$\\
$\mathcal{N}_{t}(1,1)$ & $36.48$ & $0.199$ & $0.045$ & $0.166$ & $0.027$\\
$1$ $(p=0.5)$, $10$ & $29.94$ & $0.197$ & $0.039$ & $0.182$ & $0.033$\\
$1$ $(p=0.8)$, $10$ & $19.01$ & $0.299$ & $0.092$ & $0.229$ & $0.052$\\\hline
\end{tabular}
\end{center}
\end{table}

The SPD values are close to the nominal values if the weight distribution is
rather narrow corresponding to $\tilde{\lambda}/\lambda$ close to one. Remark
that in the cases where the ratio $\tilde{\lambda}/\lambda$ is small, skewness
and excess are relatively large and correspondingly, the normal approximation
with $\gamma_{1}=\gamma_{2}=0$ is not very good. As in the limit
$n\rightarrow\infty$ both, the CPD and the SPD, approach the normal
distribution, small event numbers, or, more precisely, small values of
$\tilde{\lambda}$ are especially critical.

\section{The Poisson bootstrap}

In standard bootstrap (\cite{efronbook}) samples are drawn from the observed
observations $x_{i}$, $i=1,2,...,n$, with replacement. Poisson bootstrap is a
special re-sampling technique where to all $n$ observation $x_{i}$ Poisson
distributed numbers $n_{i}\sim\mathcal{P}_{1}(n_{i})=1/(en_{i}!)$ are
associated. More precisely, for a bootstrap sample the value $x_{i}$ is taken
$n_{i}$ times where $n_{i}$ is randomly chosen from the Poisson distribution
with mean equal to one. Samples where the sum of outcomes is different from
the observed sample size $n$, e.g. $\Sigma_{i=1}^{n}n_{i}\neq n$ are rejected.
Poisson bootstrap is completely equivalent to the standard bootstrap. It has
attractive theoretical properties \cite{babu}.

In our case the situation is different. We do not dispose of a sample of CPD
outcomes but only of a single observed value of $x$ which is accompanied by a
sample of weights. As the distribution of the number of weights is known up to
the Poisson mean, the bootstrap technique is used to infer parameters
depending on the weight distribution, To generate observations $x_{k}$, we
have to generate the numbers $n_{i}\sim$ $\mathcal{P}_{1}(n_{i})$ and form the
sum $x=\Sigma n_{i}w_{i}$. All results are kept. The resulting Poisson
bootstrap distribution (PBD) permits to estimate uncertainties of parameters
and quantiles of the CPD. Mean values derived from an infinite number of
simulated experiments and the moments extracted from the corresponding PBDs
would reproduce exactly the moments of the CPD.

\section{Applications}

In most applications we do not know the weight distribution and have to infer
it approximately from a sample of weights, $w_{i},i=1,...,n$. To this end we
replace the moments of the weight distribution by the empirical values. A
general approach to approximate the distribution of a sample starting from the
cumulants is to apply the Edgeworths \cite{gramcharlier,edgeworth} series.
Since this method is involved and not directly related to the Poisson
distribution, it has not been investigated. The Gram-Charlier series B
\cite{gramcharlier} contains explicitly a Poisson term, but it is not clear
how well the truncated series approximates the CPD. Furthermore the higher
empirical cumulants $\kappa_{3},\kappa_{4},...$ in most applications suffer
from rather large statistical fluctuations. Therefore it is often more precise
to use the values tied to the mean and the variance in the approximation by
the SPD. In addition to the SPD, we consider the simple normal approximation
and Poisson bootstrap.

\subsection{Parameter estimation from distorted measurements}

An important application of the statistics of weighted events is parameter
estimation in experiments where the data are distorted by resolution effects
\cite{bohm2012}. Typically, an experimental histogram with $m_{j}$ entries in
bin $j$ has to be compared to a theoretical prediction $x_{j}(\theta)$
depending on one or several parameters $\theta$. The prediction is obtained
from a Monte Carlo simulation which reproduces the experimental conditions and
especially the smearing by resolution effects. The variation of the prediction
with the parameter cannot be implemented by repeating the complete simulation
for each selected parameter. Therefore the simulated data which are generated
with the parameter $\theta_{0}$ according to the p.d.f. $f(\theta_{0})$ are
re-weighted by the ratio $w=$ $f(\theta)/f(\theta_{0})$. The prediction for a
histogram bin $j$ is then $x_{j}=\Sigma_{i=1}^{n_{j}}w_{ji}$ for $n_{j}$
generated events in bin $j$. To perform a least square fit of $\theta$ to a
histogram with $B$ bins, we form a $\chi^{2}$ expression where we compare
Poisson numbers $m_{j}$ times a known normalization constant $c$ to compound
Poisson numbers $x_{j}$.%
\begin{align*}
\chi^{2}  &  =%
{\displaystyle\sum\limits_{j=1}^{B}}
\frac{(cm_{j}-x_{j}(\theta))^{2}}{\delta_{j}^{2}}\;,\\
&  =%
{\displaystyle\sum\limits_{j=1}^{B}}
\frac{\left(  cm_{j}-%
{\displaystyle\sum\limits_{i=1}^{n_{j}}}
w_{ji}\right)  ^{2}}{\delta_{j}^{2}}\;.
\end{align*}
Here $\delta_{j}^{2}$ is the expected value of the numerator under the
hypothesis that the two summands in the bracket have the same expected value
$\mu$. To estimate $\delta_{j}^{2}$ first $\mu$ has to be estimated.

In the normal approximation, we compute the weighted mean of the two summands
(We suppress the index $j$.):%
\begin{equation}
\hat{\mu}_{N}=\left(  \frac{cm}{c^{2}m}+\frac{\Sigma w_{i}}{\Sigma w_{i}^{2}%
}\right)  /\left(  \frac{1}{c^{2}m}+\frac{1}{\Sigma w_{i}^{2}}\right)  \;.
\label{lambdanornal}%
\end{equation}

In the approximation based on the SPD, the value of $\mu$ can be estimated
from an approximated likelihood expression.\ The log likelihood is
\cite{bohm2012}%
\begin{equation}
\ln L(\mu)=m\ln\frac{\mu}{c}-\frac{\mu}{c}+\tilde{n}\ln\tilde{\lambda}%
-\tilde{\lambda}+const. \label{loglikspd}%
\end{equation}
where it is assumed that $m$ follows a Poisson distribution with mean $\mu/c$
and $\tilde{n}=x/s$ a Poisson distribution with mean $\tilde{\lambda}=\mu/s$,
see (\ref{enoe}), (\ref{scale}). The maximum likelihood estimate is
\cite{bohm2012}%
\begin{equation}
\hat{\mu}_{SPD}=cs\frac{\tilde{n}+m}{c+s} \label{lambdaspd}%
\end{equation}

and the corresponding estimate of $\delta^{2}$ is%
\[
\hat{\delta}_{SPD}^{2}=cs(\tilde{n}+m)\;.
\]

To evaluate the quality of the two approximations, $1000000$ experiments have
been simulated for different combinations of event numbers and weight
distributions. The results are summarized in Table 2. Here $\lambda_{n}$ and
$\lambda_{m}$ are the expected numbers of data and Monte Carlo events, $\mu$
is the mean value of $x$ that has been used in the simulation and that should
be reproduced by the estimates, $\hat{\mu}_{SPD}$ is the mean value of the SPD
estimates for $\mu$, $\sigma_{SPD}$ is the standard deviation of the estimates
and $\hat{\mu}_{N},\sigma_{N}$ are the corresponding values for the normal
approximation.\ The notation of the weight distributions is the same as above.
All estimates of $\mu$ are negatively biased but as expected the SPD values
are considerably closer to the nominal values than those of the normal
approximation. The bias for \ the SPD is in all cases below $3\%$ which is
certainly adequate for the estimation of the uncertainty $\delta$. The
fluctuations are anyway much larger than the biases both for the SPD and the
normal distribution. Both approximations are adequate, the approximation with
the SPD is slightly superior to that with the normal approximation and leads
to a simple result of the estimate of the variance $\hat{\delta}^{2}$.
Independent of the weight distribution the biases decrease with increasing
number of events. In most cases it will be possible to generate a sufficient
number of events such that $\tilde{\lambda}$ is of the order of $50$ or
larger. \begin{table}[ptb]
\caption{Comparison of the SPD and the normal approximations}%
\label{testpv}
\begin{center}
\par%
\begin{tabular}
[c]{|l|l|l|l|l|l|l|l|l|l|}\hline
$\lambda_{n}$ & $\lambda_{m}$ & weight & $\mu$ & $\hat{\mu}_{SPD}$ &
$\sigma_{SPD}$ & $\hat{\mu}_{N}$ & $\sigma_{N}$ & $\frac{\hat{\mu}_{SPD}-\mu
}{\mu}$ & $\frac{\hat{\mu}_{N}-\mu}{\mu}$\\\hline
$20$ & $20$ & $exp(-x)$ & $20$ & $19.98$ & $3.68$ & $19.10$ & $3.84$ & $0.001$
& $0.045$\\
$10$ & $10$ & $exp(-x)$ & $10$ & $9.73$ & $2.64$ & $9.11$ & $2.81$ & $0.027$ &
$0.089$\\
$10$ & $50$ & $exp(-x)$ & $10$ & $9.88$ & $1.38$ & $9.58$ & $1.59$ & $0.012$ &
$0.042$\\
$20$ & $50$ & $exp(-x)$ & $20$ & $19.84$ & $2.61$ & $19.46$ & $2.74$ & $0.008$
& $0.027$\\
$50$ & $50$ & $exp(-x)$ & $50$ & $49.78$ & $5.79$ & $49.12$ & $5.91$ & $0.004$
& $0.013$\\
$10$ & $10$ & $\mathcal{N}_{t}(1,1)$ & $12.88$ & $12.78$ & $3.13$ & $12.05$ &
$3.30$ & $0.008$ & $0.068$\\
$20$ & $20$ & $\mathcal{N}_{t}\mathcal{(}1,1)$ & $25.75$ & $25.67$ & $4.40$ &
$24.93$ & $4.53$ & $0.003$ & $0.032$\\
$20$ & $50$ & $\mathcal{N}_{t}(1,1)$ & $25.75$ & $25.69$ & $3.22$ & $25.27$ &
$3.31$ & $0.002$ & $0.019$\\
$10$ & $10$ & $u[2,3]$ & $25.00$ & $25.00$ & $5.61$ & $23.74$ & $5.87$ &
$0.000$ & $0.050$\\
$20$ & $20$ & $u[2,3]$ & $50.00$ & $50.00$ & $7.94$ & $48.74$ & $8.13$ &
$0.000$ & $0.025$\\
$50$ & $50$ & $u[2,3]$ & $125.00$ & $125.01$ & $12.54$ & $123.75$ & $12.67$ &
$0.000$ & $0.010$\\\hline
\end{tabular}
\end{center}
\end{table}

\subsection{Approximate confidence limits}

In searches for rare events frequently the identification is not unique and to
each event is attributed a weight which corresponds to the probability to be
correctly assigned. The underlying weight distribution is not known. Of
interest is the number of produced events $x=\Sigma w_{i}$ and confidence
limits for this number. The limits can be computed from the Poisson bootstrap distribution.

\begin{figure}
[ptb]
\begin{center}
\includegraphics[
trim=0.000000in 0.183834in 0.000000in 0.092168in,
height=3.0411in,
width=5.892in
]%
{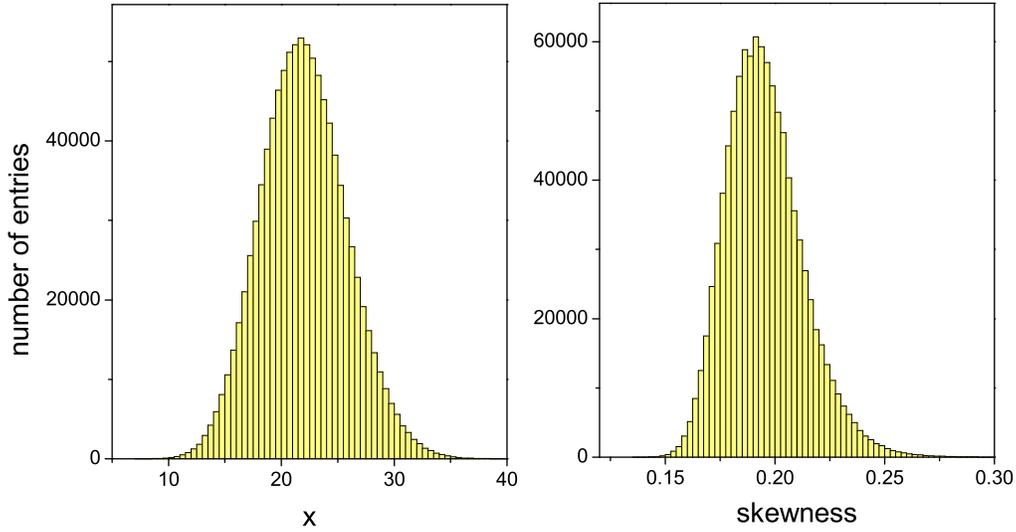}%
\caption{{\small Bootstrap distribution of the random variable }%
$x${\small (left hand)and of }$\gamma_{1}${\small (right hand).}}%
\label{bootconfi}%
\end{center}
\end{figure}

As an example, a sample of $n$ weights, with $n\sim\mathcal{P}_{50}(n)$ has
been generated with a uniform weight distribution in the interval $[0,1]$. The
value $x_{obs}=\Sigma_{i=1}^{n}w_{i}=22.01$ was obtained. The frequency plot
of the corresponding bootstrap distribution $f(x)$ is displayed in Fig.
\ref{bootconfi} left hand side. This distribution was used to derive the error
and confidence limits presented in Table \ref{confidence}. The limits
corresponding to the $\alpha$ quantiles $x$, defined by $\alpha=F(x)=\int
_{0}^{x}f(x^{\prime})dx^{\prime}$, indicated in the top line of the table, are
quoted in the second line. The usual standard error interval is $18.2<x<25.8$.

Classical confidence intervals with exact coverage cannot be computed as the
full CPD is known only approximately. But as we know the type of the
distribution for the number of events, we can improve the coverage in the
following way: We change the Poisson distribution used to generate the
bootstrap samples from $\mathcal{P}_{1}$ to $\mathcal{P}_{\mu}$ such that the
fraction of \ outcomes $x$ below $x_{obs}\mathcal{\ }$is equal to $\alpha$.
The upper limit is then $x_{up}=$ $x_{obs}\times\mu$. In a similar way the
lower limit $x_{low}$ is obtained. The central interval $x_{low}<x<x_{up}$
should then contain the unknown true value with confidence $1-2\alpha$. In the
limit where all weights are equal and the number of bootstrap samples tends to
infinity the interval would cover exactly. The obtained limits are contained
in the third line of the table. The two procedures lead to very similar
values. The modified error inteval is now $18.5<x<26.2$. As expected from the
properties of the Poisson distribution, the intervals with improved coverage
are shifted to higher values.

\begin{table}[ptb]
\caption{Confidence limits}%
\label{confidence}
\begin{center}%
\begin{tabular}
[c]{|l|l|l|l|l|l|l|l|l|}\hline
$\alpha$ & 0.01 & 0.05 & 0.10 & 0.1585 & 0.8415 & 0.90 & 0.95 & 0.99\\\hline
PBD & 13.8 & 16.0 & 17.2 & 18.2 & 25.8 & 26.9 & 28.5 & 31.4\\
PBD* & 14.4 & 16.5 & 17.6 & 18.5 & 26.2 & 27.3 & 28.9 & 32.1\\\hline
\end{tabular}
\end{center}
\end{table}

The Poisson bootstrap can be used to estimate distributions of all kinds of
parameters of the distribution. As an example the distribution of the skewness
derived from the observed weight sample is presented in the right hand plot of
Fig. \ref{bootconfi}.%

\section{Summary}

The sum of random weights where the number of weights is Poisson distributed
is described by a compound Poisson distribution. Properties of the CPD are
reviewed. The CPD is relevant for the analysis of weighted events that has to
be performed in various physics applications.

It is shown that with increasing number of events the distribution of the sum
can be approximated by a scaled Poisson distribution which coincides with the
CPD in the limit where all weights are equal. Contrary to the normal
distribution it approximately reproduces also the higher the moments of the
CPD. The SPD can be applied to the parameter estimation in situations where
the data are distorted by resolution effects. The formalism with the SPD is
simpler than that with the normal approximation and the results are more
precise. This has been demonstrated for examples with various weight distributions.

A special bootstrap method is presented which can be used to estimate from
experimental samples parameters of the underlying CPD. An example shows how it
can be applied to the estimation of confidence limits.

\section{Appendix: Proof of the Inequalities (\ref{rg1}) and (\ref{rg2})}

We apply H\"{o}lders inequality,
\[
\sum_{i}a_{i}b_{i}\leq\left(  \sum_{i}a_{i}^{p}\right)  ^{1/p}\left(  \sum
_{i}b_{i}^{p/(p-1)}\right)  ^{(p-1)/p}\;,
\]
where $a_{i},b_{i}$ are non-negative and $p>1$. For $p=2$ we obtain the
Cauchy--Schwartz inequality. Setting $a_{i}=w_{i}^{3/2}$, respectively
$b_{i}=w_{i}^{1/2}$, we get immediately the relation (\ref{rg1}) for the
skewness:
\[
\left(  \sum_{i}w_{i}^{2}\right)  ^{2}\leq\sum_{i}w_{i}^{3}\sum_{i}w_{i}\;.
\]
More generally, with $p=n-1$ and $a_{i}=w_{i}^{n/(n-1)}$, $b_{i}%
=w_{i}^{(n-2)/(n-1)}$, the inequality becomes
\[
\left(  \sum_{i}w_{i}^{2}\right)  ^{n-1}\leq\sum_{i}w_{i}^{n}\left(  \sum
_{i}w_{i}\right)  ^{n-2}\;.
\]
This formula includes also the relation (\ref{rg2}) for $n=4$.

\end{document}